\pdfoutput=1

\documentclass[sigplan,10pt]{acmart}

\setcopyright{none}             %% For review submission

\makeatletter
\fancypagestyle{firstpagestyle}{%
  \fancyhf{}%
  \renewcommand{\headrulewidth}{\z@}%
  \renewcommand{\footrulewidth}{\z@}%
    \fancyhead[L]{}%
    \fancyhead[R]{}%
}
\fancypagestyle{standardpagestyle}{%
  \fancyhf{}%
  \renewcommand{\headrulewidth}{\z@}%
  \renewcommand{\footrulewidth}{\z@}%
    \fancyhead[LE]{\@headfootfont\thepage}%
    \fancyhead[RO]{\@headfootfont\thepage}%
    \fancyhead[RE]{\@headfootfont\@shortauthors}%
    \fancyhead[LO]{\@headfootfont\shorttitle}%
    \fancyfoot[RO,LE]{}%
}
\def\@mkbibcitation{}
\makeatother

\renewcommand\footnotetextcopyrightpermission[1]{}

\usepackage[justification=centering]{caption}
\usepackage{afterpage}
\usepackage{color}
\usepackage{amssymb,amsmath,amsfonts}
\usepackage{extarrows}
\usepackage{version}
\usepackage{xspace}
\usepackage{graphicx}
\usepackage{listings}
\usepackage{wrapfig}
\usepackage{ifpdf}
\usepackage{semantic}
\usepackage{mathpartir} % this takes over some things from semantic
\usepackage{natbib}
\setcitestyle{numbers}
\usepackage{amsthm}
\usepackage{stmaryrd}
\usepackage{subfigure}
\usepackage{multirow}
\usepackage{proof}
\definecolor{darkblue}{rgb}{0.0,0.0,0.3}

\usepackage{color}
\usepackage{listings}
\usepackage[colorinlistoftodos]{todonotes}
\usepackage{tikz}
\usetikzlibrary{positioning,shadows,arrows,calc,backgrounds,fit,shapes,shapes.multipart,decorations.pathreplacing,shapes.misc,patterns}
\usepackage{tikzscale}
\usepackage{xspace}

\usepackage{hyperref}
\hypersetup{breaklinks=true}

\newcommand{\cmp}[1]{#1\hspace{-0.35em}\downarrow}

\newcommand*{\ETAL}{et al.\xspace}

\begin{document}

\title{Software Fault Isolation for Robust Compilation}

\author{Ana Nora Evans}
\authornote{
Advised by Professor Mary Lou Soffa.
University of Virginia, 85 Engineer's Way, Charlottesville, VA 22904.
Part of this work was performed as a visiting PhD student at Inria, Paris during the Summer of 2017, under the guidance of C\u{a}t\u{a}lin Hri\c{t}cu and Marco Stronati.}
\email{AnaNEvans@virginia.edu}
\affiliation{%
  \institution{University of Virginia}
}
\maketitle

Memory corruption vulnerabilities are endemic to unsafe languages, such as C, and they can even be found in safe languages that themselves are implemented in unsafe languages or linked with libraries implemented in unsafe languages. \emph{Robust compilation} mitigates the threat of linking with memory-unsafe libraries. The source language is a C-like language, enriched with a notion of a component which encapsulates data and code, exposing functionality through well-defined interfaces. Robust compilation defines what security properties a component still has, even, if one or more components are compromised. 
The main contribution of this work is to demonstrate that the compartmentalization necessary for a compiler that has the robust compilation property can be realized on a basic RISC processor using software fault isolation.

\section{Problem and Motivation}
\label{sec:intro}

Formal definitions of secure compilation have been proposed by Juglaret \ETAL \cite{DBLP:conf/csfw/JuglaretHAEP16} and, more recently, by Garg \ETAL \cite{GargHPSS17}. This work is part of the effort to propose a new definition for robust compilation of unsafe low-level languages \cite{FachiniHSELAPT17}. 
A compiler has the robust compilation property if any attack on a compiled variant of a program (a set of components) that can be mounted by a component linked and executed with it, can also be mounted at the source level by a component. In the source level semantics, it is impossible to write in another's component memory and only procedures exported by the callee and imported by the caller can be called.
Thus, for the robust compilation property to hold, a strong machine-level separation of the compiled program and the target context is necessary. Juglaret \ETAL 's \cite{JuglaretHAPST15} implementation targeted a micro-policy architecture \cite{micropolicies2015} with special tagging capabilities at the level of memory location. This work focuses on supporting the new definition of secure compilation on a generic RISC processor, without specialized hardware. We use software fault isolation \cite{Wahbe} mechanisms to provide a proof-of-concept implementation of a compiler back-end to a basic RISC machine. 

\section{Background and Related Work}

Software fault isolation was proposed in 1993 by Wahbe \ETAL \cite{Wahbe}. A distrusted module is sandboxed into its own fault domain, a logical region of the address space. To prevent it from modifying data or executing code belonging to the rest of the application, its object code is instrumented. The physical address is split logically into a segment id and offset, and the introduced instrumentation does not allow writes outside the data domain and execution to escape the code domain, other than predefined exit points. Many applications that use software fault isolation followed.  Google's Native Client \cite{Yee} uses software fault isolation to sandbox C/C++ code in the Chrome web browser. Morrissett \ETAL \cite{RockSalt} proposed a semantics of the x86 architecture and constructed a machine verified checker of Native Client. ARMor \cite{ARMor} is a machine verified system that uses software isolation to sandbox application code running on embedded processors. In this research, we combine ideas from this previous work and apply them to support robust compilation on a processor without specialized hardware.

Abadi \cite{Abadi99} defined full abstraction as the property of a compiler to preserve and reflect observational equivalence.
Achieving observational equivalence in the presence of side channels such as timing, is impossible. Instead, robust compilation focuses on only mapping back to the source level a context that induces a certain behavior on a program.
The robust compilation property for unsafe languages proposed by Fachini \ETAL \cite{FachiniHSELAPT17} is: 
\[
\forall P~C_T~t.~
  C_T \bowtie (\cmp{P}) \Downarrow t \Rightarrow  \exists C_S~t'.~ C_S \bowtie P \Downarrow t'
  \wedge t' \preccurlyeq_P t
\]

That is, for all source-level programs $P$ and all low-level contexts $C_T$ there exists a source-level context $C_S$, with no undefined behavior, such that the low-level trace $t$ of compiled $P$ linked with $C_T$ and source-level trace $t'$ of $P$ linked with $C_S$, match up to an undefined behavior in $P$.

\section{Approach and Uniqueness}

The work presented in this abstract is part of a project \citep{FachiniHSELAPT17} that aims at defining a new security property that implements a proof-of-concept compiler from a C-like language with components to two target machines: a generic RISC processor and a micro-policy machine \citep{micropolicies2015}. 
%System calls and dynamic loading are not supported yet. 
The generated executable runs on the bare hardware with the back-end compiler phase targeting the generic RISC processor. While promising, the micro-policy machine \citep{micropolicies2015} does not exist yet. Here we target a generic load-store machine with no specialized hardware for protection. The novelty of this new software fault isolation implementation is that instead of protecting an application from one or more potentially malicious libraries, all components are potentially malicious and, thus, mutually distrustful.

In our approach, a source-level program is translated from the C-like language with components to an intermediate level language that uses a similar memory model to CompCert \cite{Leroy06} enriched with a notion of component and interfaces between components. The addresses are not resolved and the interface calls between components are abstract. Our work implements a compiler pass in Coq \cite{Coq}. It takes this intermediate program and generates a RISC assembly program that satisfies the following invariants:
\begin{enumerate}
\item a component can write only within its own data memory;
\item a component can only jump within its own code memory, except for predefined exit points allowed by the interface; and
\item if after a call to another component, the execution is transferred back to the callee component, then it will always return to the instruction after the call.
\end{enumerate}

The assumptions in this research are that the basic RISC machine has a minimal load-store instruction set. The register file contains a set of registers dedicated to the software fault isolation instrumentation. The memory is unbounded and it is split into slots. The slots are allocated statically to each component and their type, code or data, is also statically determined. A physical address is an unbounded integer, with the bits starting from the least significant: offset with slot, component identifier, slot identifier. The offset and component are bounded, and the slot identifier is not. Thus, each component has an unbounded memory, but a limit on the contiguous memory it can allocate.

To enforce the first two invariants, this work uses a strategy from Wahbe \ETAL \cite{Wahbe} that has two extra instructions and three dedicated registers. Using binary bitwise operations on an address, the bits corresponding to the component identifier are set to the current one. All the data slots are odd and the instrumentation for the store instruction sets the least significant bit of the 
slot. All the code slots are even and the instrumentation for jump resets the least significant bit of the slot. Thus, no writes are possible in the code segment.

For the enforcement of the cross-component control flow, we use a dedicated protected control stack and a dedicated register for the stack pointer. The protected control stack is kept in a reserved memory, which can be accessed only from special instrumentation sequences. To ensure continuous execution of a certain number of instructions needed for managing the protected control stack, we align the instructions \cite{RockSalt}. 

The first two sandboxing invariants do not protect the current executing component, but rather protect all other components from it. Special care must be taken to protect the control stack. First, the procedures called externally are placed at an unaligned address and are preceded by a $Halt$ instruction. Thus spurious pushes onto the protected control stack are avoided. Second, to avoid the error of popping from an empty stack the execution starts with pushing the address of a $Halt$ instruction on the protected control stack and, then the execution is transferred to the main function.

\section{Results and Contributions}

The project is implemented in Coq \cite{Coq} and uses the QuickChick \cite{QuickChick} framework to test the three invariants. A test consists of the following steps: randomly generate intermediate program using QuickChick's primitives \cite{QuickChick18}, compilate with our proof-of-concept compiler, execute in simulator with recording of a log specific to each invariant using a state monad, and verify the log by a checker \cite{QuickChick18}. The intermediate programs were syntactically correct and no tests were discarded. 
Currently, we are working on simulating an attack by randomly injecting a change to the data memory of a component.

\begin{figure}[h!]
\centering
\includegraphics[width=1.0\linewidth]{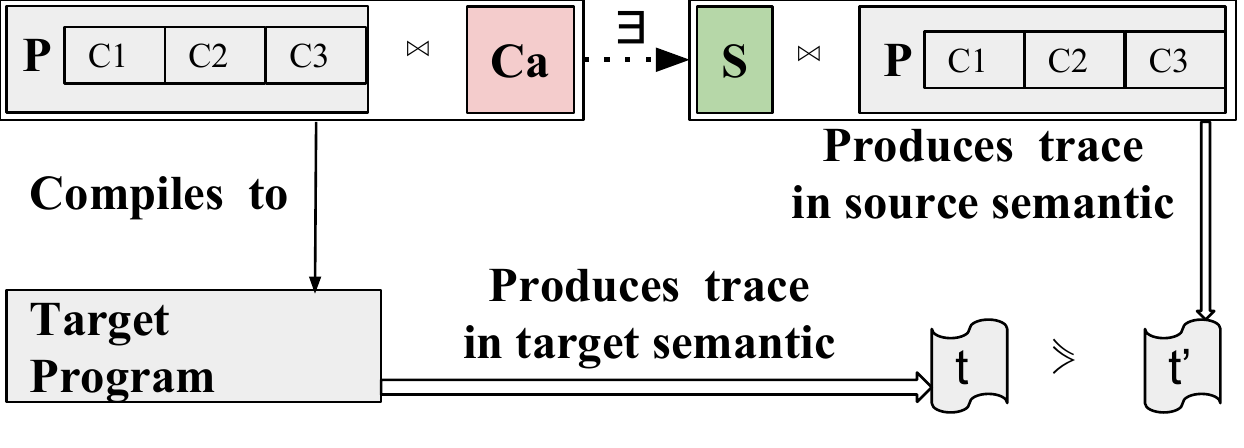}
\caption{Robust Compilation Intermediate to Target}
\label{fig:rc-def}
\end{figure}

The robust compilation property definition cannot be directly applied at the the target level, where the addresses are resolved and a certain layout in memory and instrumentation are expected. Here, the adversarial context is linked and compiled together with the program and the robust compilation property is defined as: 
\begin{equation}
\forall P~C_a~t.~
  (\cmp{(C_a \bowtie P)}) \Downarrow t \Rightarrow  \exists S~t'.~ S \bowtie P \Downarrow t'
  \wedge t' \preccurlyeq_P t
\end{equation}
In figure \ref{fig:rc-def} the program $P$ has three components, and it's linked with the adversarial component $C_a$. Together, they are compiled and executed in the target machine semantic and produce the trace $t$. By robust compilation, there exists a component $S$, with no undefined behavior, such that: $S$ together with $P$ can be executed in the intermediate semantic, producing a trace $t'$. The trace $t'$ is a prefix of trace $t$ until $S$ induces and undefined behavior in $P$.

In conclusion, we designed and implemented a compiler transformation from a RISC-like intermediate language to a basic RISC assembly language that uses software fault isolation mechanisms to provide the memory and control flow separation required by the robust compilation property. We tested the implementation using property based testing \cite{QuickCheck} and the QuickChick framework \cite{QuickChick}.

The robust compilation property does not require specialized hardware. More work is needed to support system calls and dynamic loading, but this is an encouraging first step. 

%\balance
\bibliographystyle{plainnat}
\bibliography{paper}

\end{document}